\input amstex
\documentstyle{amsppt}
\magnification=1200
\pageheight{9.5 true in}
\loadbold
\NoBlackBoxes

\def\T{\Bbb T}
\def\b{\beta}
\def\l{\lambda}
\def\G{\Gamma}
\def\z{\zeta}
\def\tht{\thetag}
\def\R{\Bbb R}
\def\C{\Bbb C}

\def\N{\Bbb N}

\def\H{\bold H}
\def\bG{{\boldsymbol\Gamma}}
\def\bGr{\bG(r,s)}
\def\O{\bold O}
\def\Hr{\H(r,s;N)}
\def\t{\theta}
\def\s{{\ssize \square}}
\def\fZ{\frak Z}
\def\fo{\frak o}
\def\fl{\frak l}
\def\fL{\frak L}

\def\Pm{P^*_\mu}

\def\wh#1{\widetilde{#1}}

\def\Fh#1#2#3{\wh{F} \left(
\matrix #1 \\ #2\endmatrix ; #3 \right)}

\def\nq{\left\lceil\frac nq \right\rceil}

\def\at{\,{\ssize\@}\,}
 
\def\sf#1#2{\left(#1\right)_{#2}}

\def\li #1 #2 #3 {\vrule height #1 pt width #2 pt depth #3 pt}

\def\hl #1 #2 #3 #4 {\rlap{ \kern #1 pt \raise -#2 pt \hbox{ \li #4 #3 0 }}}

\def\vl #1 #2 #3 #4 {\rlap{ \kern #1 pt \raise -#2 pt \hbox{ \li #3 #4 0 }}}

\def\wr #1 #2 #3 {\rlap{ \kern #1 pt \raise -#2 pt \hbox{\tenrm #3 }}}

\def\picture #1 #2 #3 \endpicture
   { \bigskip \centerline{\hbox to #1 pt
   {\nullfont #3 \hss}} \nobreak \bigskip \centerline{#2} \bigskip}

\topmatter
\title
On n-point correlations in the  log-gas at
rational temperature
\endtitle
\author
Andrei Okounkov
\endauthor
\abstract
We obtain a Ha type formula for $n$-point correlations
in the log-gas at rational temperature (or, equivalently,
$n$-point fixed time ground state correlations in the
quantum Calogero-Sutherland model for rational coupling
constant).
\endabstract
\address
University of Chicago, Dept.\ of Mathematics,
5734 S.~University Ave., Chicago, IL 60637-1546
\endaddress
\email
okounkov\@math.uchicago.edu
\endemail
\endtopmatter

\document
Let $\T^N$ denote  the $N$-dimensional torus with coordinates
$$
z_1,\dots,z_N\,, \quad z_i\in \C\,,\quad |z_i|=1 \,.
$$
Consider the following log-gas Hamiltonian
$$
\Cal H(z)=-2\sum_{i<j} \log |z_i- z_j|\,,
$$
which differs from the more standard one  by the factor of $2$. The
Gibbs state with inverse temperature $\b$ is the 
following probability measure on $\T^N$
$$
\frac1{Z(\b;N)} \prod_{i<j} |z_i-z_j|^{2\b} 
\prod \frac{dz_j}{2\pi i z_j}\,,
$$
where the partition function $Z(\b;N)$
equals (see, for example, \cite{Me})
$$
Z(\b;N)=\frac{\G(1+N\b)}{\G(1+\b)^N} \,.
$$
The same measure is the ground-state amplitude in the
quantum Calogero-Sutherland model \cite{C, Su}.
 
In this paper we suppose that $\b$ is rational 
$$
\b=p/q, \quad p,q\in\N, \quad \gcd(p,q)=1\,.
$$

Consider the $n$-point correlation functions
in the system of $(n+N)$ particles.  It is the
following measure on $\T^n$
$$
dR(n;N)=
\frac{(N+n)!}{N!}\frac{Z(\b;N)}{Z(\b;n+N)} \,
I(\z_1,\dots,\z_n;N;\b)
\prod_{i<j\le n} |\z_i - \z_j|^{2\b} 
\prod_1^n \frac{d\z_j}{2\pi i \z_j} \,, 
$$
where $I(\z_1,\dots,\z_n;N;\b)$ is the following integral
$$
I(\z_1,\dots,\z_n;N;\b):=
\frac 1{Z(\b;N)}
\int_{\T^N}  
\prod_{i\le n,j\le N} |\z_i - z_j|^{2\b}
\prod_{i<j\le N} |z_i - z_j|^{2\b} 
\prod_1^N \frac{dz_j}{2\pi i z_j} \,. \tag 1.1
$$
We are interested in the limit of $dR(n,N)$ as
$$
N\to\infty \quad\text{and}\quad \z_j\sim 1+i\,\frac{y_j}N\,, \quad
y_j\in\R\,.
$$
Since in this limit 
$$
\alignat2
&\prod_1^n \frac{d\z_j}{2\pi i \z_j} &&\sim
\frac1{(2\pi N)^n} \, dy_1 \dots dy_n\,, \\
&(n+N)!/N! &&\sim N^n\,,\\
&\prod_{i<j\le n} |\z_i - \z_j|^{2\b} &&\sim \frac1{N^{\b n (n-1)}}
\prod_{i<j\le n} |y_i - y_j|^{2\b}\,,\\
&\frac{Z(\b;N)}{Z(\b;n+N)}&&\sim \frac{\G(1+\b)^n}{(N\b)^{\b n}}  
\endalignat
$$
we have
$$
dR(n;N) \sim \frac{\G(1+\b)^n}{(2\pi)^n \b^{n\b} N^{\b n^2}} \,
I(\z_1,\dots,\z_n;N;\b) \prod_{i<j\le n} |y_i - y_j|^{2\b} \,
dy_1 \dots dy_n \,.
$$
Set
$$
I(y;\infty;\b):=\lim_{N\to\infty} \frac1{N^{\b n^2}} \,
I(\z_1,\dots,\z_n;N;\b)\,.
$$
Now we can state our {\bf main result}:

{\it 
The limit $I(y;\infty;\b)$ is the following sum of
Selberg-type integrals 
$$
\multline
I(y;\infty;\b)=\\
\sum_{\tsize \nq\le l \le n} C^n_{lp,lq-n}(\b) 
\int_{[0,1]^{lp}} dx  
\int_{[0,\infty)^{lq-n}}  d\xi \,
\,G(x\,|\,\xi;\b) 
\left|\Fh{x\,|\,\xi}{iy}{\tsize \frac1\b}\right|^2\,,
\endmultline \tag 1.2
$$
where 
$$
\multline 
G(x_1,\dots,x_r\,|\,\xi_1,\dots,\xi_s;\b):= 
\\ 
\frac{V(x)^{\frac2\b}\, V(\xi)^{2\b}}{r!s!}
\prod_{i=1}^r \left(x_i(1-x_i)\right)^{\frac1\b-1}
\prod_{j=1}^s \left(\xi_i\left(1+\frac{\xi_i}\b\right)\right)^{\b-1} 
\prod_{i=1}^r \prod_{j=1}^s \left(x_i+\frac{\xi_i}\b\right)^{-2} \,, 
\endmultline \tag 1.3
$$
$C^n_{r,s}(\b)$ is a product of $\G$-functions defined in \tht{4.8},
$V(x)$ is the Vandermonde determinant
$V(x)=\prod_{i<j}(x_i-x_j)$, 
and $\Fh{x\,|\,\xi}{iy}{\frac1\b}$ is renormalized
multivariate Bessel function \tht{2.17}.
}

The form factor \tht{1.3} is exactly the form factor
for one-particle dynamical
density-density correlations in the
Calogero-Sutherland model as 
conjectured by Haldane  and 
obtained by Ha (see \cite{Ha}, formula \tht{5.9}, and  also \cite{LPS}).
The numbers $r$ and $s$ are called in that context 
numbers of {\it quasi-holes}
and {\it quasi-particles} respectively. 

The main difference between our formula and that of  
Ha is that several different  numbers of quasi-particles
and quasi-holes contribute to correlations. 

It follows from \tht{2.17} that
$$
\alignat 2
\Fh{x\,|\,\xi}{iy}{\tsize \frac1\b}
&=O\left(\|y\|^{(lp-p+1)(lq-q-n+1)}\right)\,, &&\qquad
\|y\|\to 0 \,,\\
&=O\left((\|x\|+\|\xi\|)^{(lp-p+1)(lq-q-n+1)}\right)\,, 
&& \quad \|x\|+\|\xi\| \to 0
\endalignat
$$
provided $\nq < l\le n$. 
In other words, the contribution of summands with large $l$
is small for small $y$. 

It is well-known that for special values of $\b$
$$
\b=1/2,1,2
$$
the Bessel functions admit simple group-theoretical
interpretation and simple formulas. For example, the
case $\b=1$ is in detail discussed in \cite{OV}.

This note is organized as follows. 
Our method is quite straightforward and based upon
results of \cite{OO2}.
In the next section
we gather some useful facts about Jack polynomials
(ordinary and shifted) and Bessel functions. Then in section
3 we expand the integrand in \tht{1.1} in Jack polynomials
(which describe excitations in the CS model)
and integrate term-wise.
In section 4 we consider the asymptotic behavior of each
term of this expansion. Finally, in section 5 we collect
all previous results and obtain the above formula \tht{1.2}.

For integer $\beta$, different formulas for $n$-point correlation
functions were given by P.~J.~Forrester in \cite{F1,F2};
a formula for $2$-point correlation for general rational
$\beta$ was conjectured by P.~J.~Forrester in \cite{F3}.
For integer $\beta$ our integral can be 
integrated term-wise using an integration formula
due to Kadell \cite{Ka}. This, probably, could give a
connection between those results and ours.

I am grateful to Z.~N.~C.~Ha, Ya.~G.~Sinai, A.~Soshnikov,
and T.~Spencer for discussions during author's stay at 
IAS in spring of 1996 and to P.~J.~Forrester for
his remarks on the preliminary version of this paper. 

\head
2.~Jack polynomials, shifted Jack polynomials, and Bessel functions
\endhead

\subhead  2.1.~Jack polynomials
\endsubhead
The main reference on Jack polynomials are the book \cite{M} (Chapter VI,
especially section VI.10) and the paper \cite{St}. The Jack polynomials are also
known as $A_n$-series multivariate Jacobi polynomials and also
as excitations in the CS model. 

In this note we use notation and
results of \cite{OO2}.
We  use the parameter
$$
\t=1/\alpha
$$
inverse to the standard parameter $\alpha$ for Jack polynomials.
Jack symmetric polynomials $P_\l(x_1,\dots,x_n;\t)$ are
indexed by partitions $\l$.
We normalize $P_\l(x_1,\dots,x_n;\t)$ so that
$$
P_\l(x_1,\dots,x_n;\t)=x_1^{\l_1}\dots x_n^{\l_n} + \dots\,,
$$
where dots stand for lower monomials in lexicographic order.
The polynomial $P_\l(x;\t)$ is homogeneous of degree
$$
|\l|:=\sum \l_i \,.
$$ 

Jack polynomials are orthogonal with respect to the 
following scalar product   
$$
(f,g)_n := \frac1{Z(\t;n)}\int_{\T^n} f(z)\,\,\overline{g(z)}
\prod_{i<j\le n} |z_i-z_j|^{2\t} 
\prod_1^n \frac{dz_j}{2\pi i z_j}\,, \tag 2.1
$$
where $f$ and $g$ are polynomials in $n$ variables.

\subhead 2.2 Some notation
\endsubhead
Important constants, such as
$$
\|P_\l\|^2_n := 
\left(P_\l\,,\,P_\l\right)_n\,,
$$
can be expressed in the following functions of the partition
$\l$. 
Let 
$$
\s=(i,j)\in\l
$$
be a square in the diagram of $\l$.
Let $\l'$ denote the partition conjugate to $\l$.
The numbers
$$
\alignat2
a(\s):&=\l_i-j,&\qquad a'(\s):&=j-1,\\
l(\s):&=\l'_j-i,&\qquad l'(\s):&=i-1,
\endalignat
$$
are called arm-length, arm-colength, leg-length, and
leg-colength respectively. 
\picture 105 {Fig.~1}
\hl 0 0 108 1
\vl 0 64 64 1
\hl 0 60 10 0.2
\vl 10 60 5 0.2
\hl 10 55 5 0.2
\vl 15 55 5 0.2
\hl 15 50 30 0.2
\vl 45 50 5 0.2
\hl 45 45 5 0.2
\vl 50 45 10 0.2
\hl 50 35 10 0.2
\vl 60 35 5 0.2
\hl 60 30 5 0.2
\vl 65 30 15 0.2
\hl 65 15 5 0.2
\vl 70 15 5 0.2
\hl 70 10 10 0.2
\vl 80 10 5 0.2
\hl 80 5 25 0.2
\vl 105 5 5 0.2
\hl 0 20 65 0.2
\hl 0 25 65 0.2
\vl 35 50 50 0.2
\vl 30 50 50 0.2
\vl 29.6 25.4 5.8 0.8
\vl 34.6 25.4 5.8 0.8
\hl 29.6 20.4 5.8 0.8
\hl 29.6 25.4 5.8 0.8
\wr 3 18.5 {$\ssize{a'}$}
\wr 36.5 8 {$\ssize{l'}$}
\wr 57 18.5 {$\ssize{a}$}
\wr 36.5 46 {$\ssize{l}$}
\endpicture
\noindent
They count the number of squares in $\l$ which lie 
to the east, west, south, and north of $\s$ respectively
(see Fig.~1).

Consider  the following 
hook-length products:
$$
\align
H(\l;\t)&:=\prod_{\s\in\l}(a(\s)+\t\, l(\s)+1) \\
H'(\l;\t)&:=\prod_{\s\in\l}(a(\s)+\t\, l(\s)+\t)
\,.
\endalign
$$
Note that
$$
H(\l';\t)=\t^{|\l|}\,H'(\l;\t^{-1})\,,
\qquad H'(\l';\t)=\t^{|\l|}\,H(\l;\t^{-1})\,, \tag 2.2 
$$
where $\l'$ stands for the conjugate of the partition $\l$.
Introduce also the following analog of the shifted
factorial
$$
\sf{t;\t}{\l} = \prod_{\s\in\l}(t+a'(\s)-\t\, l'(\s))\,,
$$
where $t$ is a variable. 
If $\mu=(m)$ then 
$$
\sf{t;\t}{(m)}=t(t+1)\dots(t+m-1)
$$ 
is the standard shifted factorial.

We have (see \cite{M}, VI.10.37)
$$
\|P_\l\|^2_n = 
\frac{\sf{\t n;\t}{\l}}{\sf{\t(n-1)+1;\t}{\l}}
\frac{H(\l;\t)}{H'(\l;\t)}\,, \tag 2.3 
$$
and (see \cite{M}, VI.10.20) 
$$
P_\l(\underbrace{1,\dots,1}_{\text{$n$ times}};\t) 
= \sf{n\t;\t}{\l} \big/ H'(\l;\t) \,. \tag 2.4
$$

\subhead 2.3 Variations of Jack polynomials
\endsubhead
Consider the expansion of Jack polynomials $P_\l$ in polynomials
$$
p_\rho(x):= \prod_{\rho_i > 0} \sum x_j^{\rho_i} \,,
$$
where $\rho$ is a partition.
The coefficients of this expansion
$$
P_\l(x;\t) =\sum_\rho \chi^\l_\rho(\t) \, p_\rho(x) \tag 2.5
$$
are natural $\t$-analogs of characters of symmetric groups.

Using \tht{2.5} one can define the two following modification 
of Jack polynomials. 
First, it is clear that for a natural number $k$
$$
P_\l(\underbrace{x_1,\dots,x_1}_{\text{$k$ times}},
\underbrace{x_2,\dots,x_2}_{\text{$k$ times}}, \dots; \t)
=
\sum_\rho k^{\ell(\rho)}\, \chi^\l_\rho(\t) \, p_\rho(x)\,,
$$
where $\ell(\rho)$ stands for the number of parts of $\rho$.
Now, by definition,
we set for an arbitrary $k$
$$
P_\l(x\at k;\t) :=\sum_\rho 
k^{\ell(\rho)}\, \chi^\l_\rho(\t) \, p_\rho(x)\,. \tag 2.6
$$

Consider also the following super-analogs of the functions  
$p_\rho(x)$
$$
p_\rho(x\,|\,\xi;\t) := \prod_{\rho_i > 0} 
\left(\sum x_j^{\rho_i}+(-\t)^{\rho_i-1}\sum \xi_j^{\rho_i}
\right)\,,
$$
and define super-Jack polynomials (see \cite{KOO}) by
$$
P_\l(x\,|\,\xi;\t) :=\sum_\rho \chi^\l_\rho(\t) \, p_\rho(x\,|\,\xi;\t)\,. \tag 2.7
$$

\subhead 2.4 Shifted Jack polynomials
\endsubhead
Shifted Jack polynomials were studied in the papers
\cite{S1,OO,Ok1,KS,Ok2,OO2}.
The shifted Jack polynomial
$$
\Pm(x_1,\dots,x_n;\t)
$$
is the unique
polynomial satisfying the following conditions:
\roster
\item it is symmetric in variables $x_i-\t i$, $i=1,\dots,n$,
\item it has degree $|\mu|$,
\item $\Pm(\mu)=H(\mu)$,
\item $\Pm(\l)=0$ unless $\mu\subset\l$,
\endroster
where we assume that $\mu$ and $\l$ have at most $n$ parts.
Explicit formulas for shifted Jack polynomials were found in \cite{Ok2}
(see also \cite{Ok3} for more general results).

The following theorem was proved in \cite{OO2}.
\subhead 2.5 Binomial formula
\endsubhead
$$
\frac{P_\l(1+x_1,\dots,1+x_n;\t)}
{P_\l(1,\dots,1;\t)} =
\sum_{\mu\subset\l}
\frac{\Pm(\l_1,\dots,\l_n;\t)}{H(\mu)}
\frac{P_\mu(x_1,\dots,x_n;\t)}
{P_\mu(1,\dots,1;\t)} \,. \tag 2.9
$$
Recall that there is a product formula \tht{2.4} for denominators
$P_\l(1,\dots,1;\t)$ and $P_\mu(1,\dots,1;\t)$.

For $k\in\N$ we have
$$
\frac{P_\l(1+x_1,\dots,1+x_n\at k;\t)}
{P_\l(1\at kn;\t)} =
\sum_{\mu\subset\l}
\frac{\Pm(\l_1,\dots,\l_n;\t)}{H(\mu)}
\frac{P_\mu(x_1,\dots,x_n\at k;\t)}
{P_\mu(1\at kn;\t)} \,. \tag 2.10
$$
Since \tht{2.10} is an identity of rational functions in $k$
it is true identically.

We shall need the identity \tht{2.10} for 
$$
k\to \t^{-1}\,.
$$
A very important feature of this specialization 
is that the product
$$
\sf{k n \t;\t}{\l} \to \sf{n;\t}{\l}\,, \quad k\to \t^{-1}
$$
may vanish and so will the denominators in \tht{2.10}.

Denote by
$$
\fl(\s;t,\t):=t+a'(\s)-\t l'(\s)
$$
the linear functions that enters the definition of
$\sf{t;\t}{\l}$
$$
\sf{t;\t}{\l}=\prod_{\s\in\l}\fl(\s;t,\t)\,.
$$
Denote by 
$$
\fZ(t,\t):=\left\{\s,\,\fl(\s;t,\t)=0\right\}
$$
the zero set of this linear function.
For example, the squares from $\fZ(3,4)$ are filled black in Fig.~2
\picture 132 {Fig.\ 2}
\hl 0 0 132 1
\hl 0 5 132 0.2
\hl 0 10 132 0.2
\hl 0 15 132 0.2
\hl 0 20 132 0.2
\hl 0 25 132 0.2
\hl 0 30 132 0.2
\hl 0 35 132 0.2
\hl 0 40 132 0.2
\vl 0 42 42 1
\vl 5 42 42 0.2
\vl 10 42 42 0.2
\vl 15 42 42 0.2
\vl 20 42 42 0.2
\vl 25 42 42 0.2
\vl 30 42 42 0.2
\vl 35 42 42 0.2
\vl 40 42 42 0.2
\vl 45 42 42 0.2
\vl 50 42 42 0.2
\vl 55 42 42 0.2
\vl 60 42 42 0.2
\vl 65 42 42 0.2
\vl 70 42 42 0.2
\vl 75 42 42 0.2
\vl 80 42 42 0.2
\vl 85 42 42 0.2
\vl 90 42 42 0.2
\vl 95 42 42 0.2
\vl 100 42 42 0.2
\vl 105 42 42 0.2
\vl 110 42 42 0.2
\vl 115 42 42 0.2
\vl 120 42 42 0.2
\vl 125 42 42 0.2
\vl 130 42 42 0.2
\hl 5 10 5 5
\hl 25 15 5 5
\hl 45 20 5 5
\hl 65 25 5 5
\hl 85 30 5 5
\hl 105 35 5 5
\hl 125 40 5 5
\wr -50  25 {$\fZ(3,4)=$}
\endpicture
\noindent
In general, if
$$
t\in\N\,, \quad \t=q/p\,, \quad \gcd(p,q)=1
$$
the set $\fZ(t,\t)$ has the following parameterization
$$
\fZ(t,\t)=\left\{
(lp+1,lq-t+1)\,, l\in\N\,, l\ge
\left\lceil\frac tq \right\rceil \right\}\,.
$$
Set also 
$$
\fo(\l;t,\t):=\#\left\{\l\cap\fZ(t,\t)\right\}\,.
$$
Then $\fo(\l;n,\t)$ is the order of zero of the denominator
in the LHS of \tht{2.10} as $k\to 1/\t$
$$
\sf{k n \t;\t}{\l} = O\left((k-1/\t)^{\fo(\l;n,\t)}\right)\,,
\quad k\to 1/\t\,.
$$

Define the following renormalized products
$$
\wh{\sf{t;\t}{\l}} := 
\prod_{\s\in\l\setminus\fZ(t,\t) }\fl(\s;t,\t)\,,
$$
and
$$
\wh{P_\l(1\at k;\t)} := \wh{\sf{k\t;\t}{\l}} \big/ H'(\l;\t) \,. \tag 2.11
$$
Then we have the following renormalized binomial formula 
$$
\multline
\frac{P_\l(1+x_1,\dots,1+x_n\at \t^{-1};\t)}
{\wh{P_\l(1\at n/\t;\t)}} = \\
\sum\Sb\mu\subset\l\\
\fo(\mu;n,\t)=\fo(\l;n,\t)
\endSb
\frac{\Pm(\l_1,\dots,\l_n;\t)}{H(\mu)}
\frac{P_\mu(x_1,\dots,x_n\at \t^{-1};\t)}
{\wh{P_\mu(1\at n/\t;\t)}} \,. 
\endmultline \tag 2.12
$$
In plain words, the summation in the LHS of \tht{2.12}
ranges over those diagrams $\mu\subset\l$ that contain
as many elements of $\fZ(n,\t)$ (``zero squares'') as 
$\l$ does. 

\subhead
2.7~Bessel functions (\cite{D,Op,J,OO2})
\endsubhead
For a real vector $l=(l_1,\dots,l_n)$ denote by
$$
[l]:=([l_1],\dots,[l_n])
$$
its integral part. 
Suppose that $l_1\ge\dots\ge l_n$.
By definition, put
$$
F(l,y;\t) :=\lim_{N\to\infty}
\frac{P_{[N l]} (1+y_1/N,\dots,1+y_n/N;\t)}
{P_{[N l]} (1,\dots,1;\t)} \,. 
$$
The following formula follows (see \cite{OO2}) from the
binomial formula
$$
F(l,y;\t) =\sum_\mu 
\frac{ P_\mu(l_1,\dots,l_n;\t)
\, P_\mu(y_1,\dots,y_n;\t)}{H(\mu)\, P_\mu(1,\dots,1;\t)} 
\,. \tag 2.14
$$
The functions  $F(l,y;\t)$ are called the {\it Bessel functions}
because they are in the same relation to Jack polynomials as ordinary Bessel
functions to Jacobi polynomials. 

\subhead
2.8~Renormalized Bessel functions
\endsubhead
Consider the following modification of the Bessel function.   
Denote by $\bGr$ the following $\G$-shaped set 
$$
\bGr=\left\{\s=(i,j)\,, \text{ $i\le r$ or $j\le s$}\right\}\,.
$$ 
and denote by $\H(r,s)$ the set of all partitions $\l$ such
that
$$
\l\in\bGr \,.
$$
In other words, the set $\H(r,s)$ consists of those partitions
that do not contain the square $(r+1,l+1)$. An example of such
partition is given in Fig.~3
\picture 90 {Fig.~3}
\hl 0 0 90 1
\hl 20 15 70 1
\vl 0 60 60 1
\vl 20 60 45 1
\hl 20 20 5 1
\vl 25 20 5 1
\hl 0 55 5 0.2
\vl 5 55 15 0.2
\hl 5 40 5 0.2
\vl 10 40 10 0.2
\hl 10 30 5 0.2
\vl 15 30 5 0.2
\hl 15 25 5 0.2
\vl 40 15 5 0.2
\hl 40 10 20 0.2
\vl 60 10 5 0.2
\hl 60 5 25 0.2
\vl 85 5 5 0.2
\wr 27.5 25 {$\ssize{(r+1,s+1)}$}
\endpicture
\noindent 
The elements of 
$\H(r,s)$ are ``fat hooks'' with $\le r$ arms and $\le s$ legs.

Now fix some $r$ and $s$ and let 
$$
\l=\l(N)\,\in\H(r,s)\,, \quad N\to\infty\,,
$$
be a sequence of partitions such that the following
limits exist
$$
\align
x_i&:=\lim \frac{\l_i}N\,, \quad i=1,\dots,r\\
\xi_j&:=\lim \frac{\l'_j}N\,, \quad j=1,\dots,s \,.
\endalign
$$
This kind of growth of a partition was considered,
probably, for the first time in \cite{VK}.
It follows from  \cite{KOO}, Theorem 7.1, that
$$
\frac{\Pm(\l;\t)}{N^{|\mu|}} \to 
P_\mu(x_1,\dots,x_r\,|\,\xi_1,\dots,\xi_s;\t)\,,
\quad N\to\infty \,. \tag 2.15
$$
where the RHS was defined in \tht{2.7}. 
Set
$$
\fo_{n,r,s;\t}:=\#\left\{\bGr\cap\fZ(n,\t)\right\}\,.
$$
and  suppose that $x_r>0$ and $\xi_s>0$. 
Then
$$
\fo(\l;n,\t)=\fo_{n,r,s;\t}
$$
for all sufficiently large $N$.

Now we consider the following renormalized Bessel function
$$
\Fh{x_1,\dots,x_r\,|\,\xi_1,\dots,\xi_s}
{y_1,\dots,y_n}{\t} := 
\lim_{N\to\infty}
\frac{P_{\l} (1+y_1/N,\dots,1+y_n/N \at \t^{-1};\t)}
{\wh{P_{\l} (1\at n/\t;\t)}} \,. \tag 2.16
$$
It follows from \tht{2.15}, \tht{2.12} and \tht{2.11} that
$$
\multline
\Fh{x_1,\dots,x_r\,|\,\xi_1,\dots,\xi_s}
{y_1,\dots,y_n}{\t} = \\
\sum\Sb
\mu\in\H(r,s)\\
\fo(\mu;n,\t)=\fo_{n,r,s;\t} \endSb
\frac{H'(\mu;\t)}{H(\mu;\t)}
\frac{P_\mu(x\,|\,\xi;\t)
\, P_\mu(y \at \t^{-1};\t)}
{\wh{\sf{n;\t}{\mu}}} \,. 
\endmultline \tag 2.17
$$
The sum in \tht{2.17} ranges over all those $\mu\in\H(r,s)$
that have as many ``zero squares'' (squares in $\fZ(n,\t)$)
as possible.

\head
3.~Expansion of $I(\z_1,\dots,\z_n;N;\b)$
\endhead

Since $|\z_i|=|z_j|=1$ we have
$$
\prod_{i\le n,j\le N} |\z_i - z_j|^{2\b} =
\prod_{i\le n,j\le N} (1- \bar\z_i z_j)^{\b}  (1- \z_i \bar z_j)^{\b}
$$
In this section we shall expand the product
$$
\prod_{i\le n,j\le N} (1- \bar\z_i z_j)^{\b}
$$
in Jack polynomials in $z_1,\dots,z_N$ and then integrate 
in \tht{1.1} term-wise using \tht{2.3}.

Denote by $\Hr$ the subset of $\H(r,s)$ which consists of 
partitions $\l$ satisfying
$$
\l_1 \le N\,.
$$
We have the following 

\proclaim{Proposition 3.1}
$$
\multline
\prod_{i\le n,j\le N} (1- \bar\z_i z_j)^{\b}=\\
\sum_{\l\in\H(np,n(q-1);N)} (-1)^{|\l|} P_{\l'}(z_1,\dots,z_N;\b)
P_\l(\bar\z_1,\dots,\bar\z_n \at  \b;{\tsize \frac 1\b}) \,.
\endmultline 
$$
\endproclaim

Recall that
$$
\b=p/q\,.
$$
This type of expansion is standard; the only non-standard part is
the condition $\l\in\H(np,n(q-1);N)$. In the proof we shall need
the following 

\proclaim{Lemma 3.2} Let $L(r,s)$ be the linear span of polynomials
$P_\l(x;\t)$ such that  $\l\in\H(r,s)$. 
Then
\roster
\item"(a)" $L(r,s)$ is the linear span of the products of the 
form
$$
P_{(k_1)}(x;\t) \dots P_{(k_r)}(x;\t) \,
e_{m_1}(x) \dots e_{m_s}(x)\,,
$$
where $k_1,\dots,k_r,m_1,\dots,m_s$ range over all 
non-negative integers. Here $(k)$ stands for
the partition with the only part equal to $k$ and
$e_m$ is the $m$-th elementary symmetric function.
\item"(b)" $L(r_1,s_1) \cdot L(r_2,s_2) = L(r_1+r_2,s_1+s_2)$
\,.
\endroster
\endproclaim

\demo{Proof of Lemma} The (a) part follows by induction
from formulas for multiplication of a Jack
polynomial by $P_{(k)}(x;\t)$ or $e_m(x)$,
see \cite{M}, section VI.6 and section VI.10.
The (b) part follows from (a). \qed
\enddemo

\demo{Proof of Proposition}
We have (\cite{M}, VI.5.4)
$$
\prod_{i,j}(1-x_i y_j) = \sum_\l (-1)^{|\l|} 
P_{\l'}(x;\t) P_\l(y;\t^{-1}) \,.
$$
For an integer $k$ it implies
$$
\prod_{i,j}(1-x_i y_j)^k = \sum_\l (-1)^{|\l|} 
P_{\l'}(x;\t) P_\l(y\at k;\t^{-1}) \,. \tag 3.1
$$
Since both sides of \tht{3.1} are power series in $x$ and $y$
with coefficients in polynomials in $k$ we can let $k$
be any number. Therefore
$$
\prod_{i\le n,j\le N} (1- z_j \bar\z_i )^{\b}=
\sum_{\l} (-1)^{|\l|} P_{\l'}(z_1,\dots,z_N;\b)
P_\l(\bar\z_1,\dots,\bar\z_n \at  \b;{\tsize \frac 1\b}) \,. \tag 3.2
$$
Note that
$$
P_{\l'}(z_1,\dots,z_N;\b)=0
$$
if the partition $\l'$ has more than $N$ parts, that is,  if 
$$
\l_1 >  N \,.
$$
It remains to show that the sum in \tht{3.2} is over $\l\in\H(np,n(q-1))$.

First, consider the case $n=1$. Then since $P_\l$ is homogeneous
we have 
$$
P_\l(\bar\z_1\at \b;{\tsize \frac 1\b})=(\bar\z_1)^{|\l|} P_\l(1\at \b;{\tsize \frac 1\b})\,.
$$
By virtue of \tht{2.4}
$$
P_\l(1\at \b;{\tsize \frac 1\b})=\frac1{H'(\l)} \prod_{(i,j)\in\l} (j-(i-1)/\b) \,.
$$
This product vanishes if
$$
(p+1,q)\in\l\,.
$$
Therefore if $n=1$ then the sum in \tht{3.2} is over 
$\l\in\H(p,q-1)$, or in other words
$$
\prod_{j\le N} (1- \bar\z_1 z_j)^{\b} \, \in  \,L(p,q-1) \tag 3.3
$$
as a power series in $z_1, \dots,z_N$. 

For general $n$ we have
$$
\prod_{i\le n,j\le N} (1- \bar\z_i z_j)^{\b} =
\prod_{i\le n} \left(\prod_{j\le N} (1- \bar\z_i z_j)^{\b} \right)\,.
$$
Therefore by \tht{3.3} and  the part (2) of the Lemma we have
$$
\prod_{i\le n,j\le N} (1- \bar\z_i z_j)^{\b}  \, \in
\,L(np,n(q-1)) 
$$
as a power series in $z_1, \dots,z_N$. Hence the 
sum in \tht{3.2} is over $\l\in\H(np,n(q-1))$.
\qed
\enddemo

Integrating in \tht{1.1} term-wise we obtain the following

\proclaim{Corollary 3.3}
$$
\multline
I(\z_1,\dots,\z_n;N;\b)=\\
\sum_{\l\in\H(np,n(q-1);N)}
\|P_{\l'}(z;\b)\|_N^2 \,\, |P_\l(\z_1,\dots,\z_n \at  \b;{\tsize \frac 1\b})|^2
\,.
\endmultline \tag 3.4 
$$
\endproclaim

\head
4.~Asymptotics of each summand in \tht{3.4}
\endhead

In this section we consider the the asymptotic behavior
of the number 
$$
\|P_{\l'}(z;\b)\|^2_N \,\,
|P_\l(\z_1,\dots,\z_n\at \b;{\tsize \frac 1\b})|^2
\,,\quad \l\in\H(r,s)\,,
$$
as 
$$
N\to\infty\,,\quad \l\to\infty
$$ 
in such a way that the limits
$$
\alignat2
x_i&=\lim \l_i\big/N\,, \quad &&i=1,\dots,r\,, \\
\xi_j&=\lim \l'_j\big/N\,, \quad &&j=1,\dots,s\,, 
\endalignat
$$ 
exist.
Note that if $\l\in\Hr$ then 
$$
x_i\le 1\,, \quad i=1,\dots,r\,.
$$
First, from the equality
$
{\G(t+m)}/{\G(t)}=t(t+1)(t+2)\dots(t+m-1)
$, 
$m\in\N$,
one deduces the following 

\proclaim{Lemma 4} We have 
$$
\align
\sf{t;\t}{\l}&=
\prod_{i=1}^r 
\frac{\G(t+\l_i+(1-i)\t)}{\G(t+(1-i)\t)}
\prod_{j=1}^s
\t^{\l'_j-r} 
\frac{\G((t+j-1)/\t+1-r)}{\G((t+j-1)/\t+1-\l'_j)} \tag 4.1
\\
H(\l;\t)&=
\G(1/\t)^{-s}
\prod_{i=1}^r \prod_{j=1}^s
(\l_i-j+(\l'_j-i)\t+1) \tag 4.2\\
&\times
\prod_{i=1}^r \G(\l_i-s+1+(r-i)\t)
\prod_{j=1}^s  \t^{\l'_j-r}  \,\G(\l'_j-r+(s+1-i)/\t) \\
&\times 
\prod_{i=1}^{r-1} \prod_{k=i+1}^r 
\frac{\G(\l_i-\l_k+1+(k-i-1)\t)}{\G(\l_i-\l_k+1+(k-i)\t)}\\
&\times
\prod_{j=1}^{s-1} \prod_{k=j+1}^s  
\frac{\G(\l'_j-\l'_k+(k-j)/\t)}{\G(\l'_j-\l'_k+(k-j+1)/\t)} \\
%
%
H'(\l;\t)&=
\G(\t)^{-r}
\prod_{i=1}^r \prod_{j=1}^s
(\l_i-j+(\l'_j-i)\t+\t) \tag 4.3\\
&\times
\prod_{i=1}^r \G(\l_i-s+(r-i+1)\t)
\prod_{j=1}^s  \t^{\l'_j-r} \, \G(\l'_j-r+1+(s-i)/\t) \\
&\times 
\prod_{i=1}^{r-1} \prod_{k=i+1}^r 
\frac{\G(\l_i-\l_k+(k-i)\t)}{\G(\l_i-\l_k+(k-i+1)\t)}\\
&\times
\prod_{j=1}^{s-1} \prod_{k=j+1}^s  
\frac{\G(\l'_j-\l'_k+1+(k-j-1)/\t)}{\G(\l'_j-\l'_k+1+(k-j)/\t)} 
\,. 
\endalign
$$
provided $\l\in\H(r,s)$ and $\l\notin\H(r-1,s-1)$.
\endproclaim

First, consider (recall \tht{2.3})
$$
\|P_{\l'}(z;\b)\|^2_N = 
\frac{\sf{ N\b;\b}{\l'}}{\sf{1+(N-1)\b;\b}{\l'}}
\frac{H(\l';\b)}{H'(\l';\b)} \,.
$$
From the relation
$$
\frac{\G(x+a)}{\G(x)} \sim x^a\,, \quad x\to\infty
$$
and the Lemma it follows that
$$
\frac{\sf{ N\b;\b}{\l'}}{\sf{1+(N-1)\b;\b}{\l'}}
\sim
\prod_{i=1}^r (1-x_i)^{{\tsize \frac 1\b}-1}
\prod_{j=1}^s (1+\xi_i/\b)^{\b-1} 
$$
and that
$$
\frac{H(\l';\b)}{H'(\l';\b)} \sim 
\frac{\G(\b)^s}{\G({\tsize \frac 1\b})^r}
\prod_{i=1}^r \l_i^{{\tsize \frac 1\b}-1}
\prod_{j=1}^s {\l'_j}^{1-\b} \,.
$$
Altogether,
$$
\|P_{\l'}(z;\b)\|^2_N \sim
\frac{\G(\b)^s}{\G({\tsize \frac 1\b})^r}
\prod_{i=1}^r \l_i^{{\tsize \frac 1\b}-1}
\prod_{j=1}^s {\l'_j}^{1-\b}
\prod_{i=1}^r (1-x_i)^{{\tsize \frac 1\b}-1}
\prod_{j=1}^s (1+\xi_i/\b)^{\b-1}  \,. \tag 4.4
$$

Now consider
$$
|P_\l(\z_1,\dots,\z_n\at \b;{\tsize \frac 1\b})|^2
$$
Since by \tht{2.16}
$$
\frac{|P_\l(\z_1,\dots,\z_n \at  \b;{\tsize \frac 1\b})|^2}
{|\wh{P_\l(1\at  n\b;{\tsize \frac 1\b})}|^2} \to
\left|\Fh{x_1,\dots,x_r\,|\,\xi_1,\dots,\xi_s}
{i y_1,\dots,i y_n}{\tsize \frac 1\b}\right|^2 \tag 4.5 
$$
it suffices to consider the asymptotics of
$$
|\wh{P_\l(1\at  n\b;{\tsize \frac 1\b})}|^2=
\left|\wh{\sf{n;{\tsize \frac1\b}}{\l}} \Big/ 
H'(\l;{\tsize \frac1\b}) \right|^2\,.
$$
Using the obvious relation
$$
\prod_{i=1}^r 
\frac{\G(t+\l_i+(1-i)\t)}{\G(t+(1-i)\t)}
=\prod_{i=1}^r \prod_{j=1}^s (t+(j-1)-(i-1)\t)
\prod_{i=1}^r 
\frac{\G(t+\l_i+(1-i)\t)}{\G(t+s+(1-i)\t)}\,.
$$
one obtains from  the Lemma 
$$
\align
|\wh{P_\l(1\at  n\b;{\tsize \frac 1\b})}|^2 &\sim 
\G({\tsize \frac 1\b})^{2r} \,
\left(\wh{\sf{n;{\tsize \frac1\b}}{(s^r)}}\right)^2 \tag 4.7
\\
&\times 
\prod_{i=1}^r \G^{-2} (n+s+(1-i)/\b)
\prod_{j=1}^s \G^{-2} (r-(n+j-1)\b)\\
&\times
\prod_{i=1}^r \l_i^{2(n+s-r/\b)} 
\prod_{j=1}^s {\l'_j}^{2(r-(n+s)\b)+2(\b-1)}\\
&\times V(\l)^{2/\b}\, V(\l')^{2\b} \, 
\prod_{i=1}^r \prod_{j=1}^s (\l_i+\l'_i/\b)^{-2} 
\,,
\endalign
$$
where $(s^r)$ denotes the following partition
$$
(s^r)=(\underbrace{s,\dots,s}_{\text{$r$ times}})\,.
$$
Collect all $\G$-functions in one constant as follows
$$
\multline
C^n_{r,s}(\b):= 
\G({\tsize \frac 1\b})^{r}\,
\G(\b)^{s}\,\left(\wh{\sf{n;{\tsize \frac1\b}}{(s^r)}}\right)^2  \\
\times 
\prod_{i=1}^r \G^{-2} (n+s+(1-i)/\b)
\prod_{j=1}^s \G^{-2} (r-(n+j-1)\b) \,. 
\endmultline \tag 4.8 
$$
Recall that a tilde over a product means that we omit zero factors.

Consider the following linear function
$$
\fL(r,s;t,\t)=\fl((r+1,s+1);t,\t)=t+s-r\t\,.
$$
By definition, set
$$
\align
G^0_{n,r,s}(x_1,\dots,x_r\,|\,\xi_1,\dots,\xi_s;\b)=& 
\prod_{i=1}^r x_i^{2\fL(r,s;n,\frac1\beta)+({\tsize \frac 1\b}-1)}
\prod_{j=1}^s {\xi_j}^{-2\b\fL(r,s;n,\frac1\beta)+(\b-1)} \\
\times&
V(x)^{2/\b}\, V(\xi)^{2\b} \, 
\prod_{i=1}^r \prod_{j=1}^s (x_i+\xi_i/\b)^{-2}\\
\times&
\prod_{i=1}^r (1-x_i)^{{\tsize \frac 1\b}-1}
\prod_{j=1}^s (1+\xi_i/\b)^{\b-1}\,.
\endalign
$$
Note that the first two lines of the last formula
represent a homogeneous function of degree
$$
\b n^2 - r - s - \b\fL^2(r,s;n,\frac1\beta)\,.
$$
With this notation the relations \tht{4.4,4.5,4.7} imply that 
$$
\multline
\frac{C^n_{r,s}(\b)}{N^{\b n^2}}\, \|P_{\l'}(z;\b)\|^2_N \,\,
|P_\l(\z_1,\dots,\z_n\at \b;{\tsize \frac 1\b})|^2
\sim \\
\frac1{N^{r+s+\b\fL^2(r,s;n,\frac 1\b)}} \,
G^0_{n,r,s}\left(
x\,|\, \xi;\b\right) \,
\left|
\Fh{x_1,\dots,x_r\,|\, \xi_1,\dots,\xi_s}
{iy_1,\dots,iy_n}{\tsize \frac 1\b}
\right|^2 \,.
\endmultline  \tag 4.9
$$

As we shall see the asymptotically
non-zero contributions to correlations functions
arise from values of $r$ and $s$ satisfying
$$
\fL(r,s;n,{\tsize \frac 1\b})=0
$$
The other values of $r$ and $s$ get suppressed by the
factor
$$
\frac1{N^{\b\fL^2(r,s;n,\frac 1\b)}}\to 0\,, \quad N\to\infty\,.
$$

\head
5.~Asymptotics of $I(\z_1,\dots,\z_n;N;\b)$
\endhead

In this section we consider the asymptotic behavior 
of the integral 
$$
\multline 
\frac1{N^{\b n^2}}\,I(\z_1,\dots,\z_n;N;\b)=\\
\sum_{\l\in\H(np,n(q-1);N)} \frac1{N^{\b n^2}}\,
\|P_{\l'}(z;\b)\|_N^2 \,\, |P_\l(\z_1,\dots,\z_n \at \b;
{\tsize \frac 1\b})|^2 \,,
\endmultline 
$$
where in the RHS we recall the expansion \tht{3.4}.
We break this  into pieces according to the number of
elements of $\fZ(n,\t)$ (``zero squares'') in $\l$. 

For 
$$
l=\nq,\dots,n
$$
set
$$
\O(l,n;\b):=\H(lp,lq-n;N)\setminus\H((l-1)p,(l-1)q-n)\,.
$$
Since
$$
\bG(np,nq-n)\cap \fZ(n,q/p) =
\left\{(lp+1,lq-n+1)\,,l=\nq,\dots,n-1\right\}
$$ 
we have 
$$
\Big( \l\in \O(l,n;\b)\Big) \Leftrightarrow
\Big( \fo(\l;n,{\tsize \frac1\b}) = l-\nq  \Big) \,,
$$

For example, if $n=3$ and $\b=1/4$ then we have 3 pieces
which consist of diagrams containing 0,1, or 2 of the
2 possible black squares in Fig.~4
\picture 132 {Fig.\ 4}
\hl 0 0 132 1
\hl 0 5 132 0.2
\hl 0 10 132 0.2
\hl 0 15 132 0.2
\hl 45 16 87 1
\hl 0 20 45 0.2
\hl 0 25 45 0.2
\hl 0 30 45 0.2
\hl 0 35 45 0.2
\hl 0 40 45 0.2
\vl 0 42 42 1
\vl 5 42 42 0.2
\vl 10 42 42 0.2
\vl 15 42 42 0.2
\vl 20 42 42 0.2
\vl 25 42 42 0.2
\vl 30 42 42 0.2
\vl 35 42 42 0.2
\vl 40 42 42 0.2
\vl 45 42 42 0.2
\vl 45 42 27 1 
\vl 50 15 15 0.2
\vl 55 15 15 0.2
\vl 60 15 15 0.2
\vl 65 15 15 0.2
\vl 70 15 15 0.2
\vl 75 15 15 0.2
\vl 80 15 15 0.2
\vl 85 15 15 0.2
\vl 90 15 15 0.2
\vl 95 15 15 0.2
\vl 100 15 15 0.2
\vl 105 15 15 0.2
\vl 110 15 15 0.2
\vl 115 15 15 0.2
\vl 120 15 15 0.2
\vl 125 15 15 0.2
\vl 130 15 15 0.2
\hl 5 10 5 5
\hl 25 15 5 5
\hl 45 20 5 5
\endpicture

We write
$$
\multline 
\frac1{N^{\b n^2}}\,I(\z_1,\dots,\z_n;N;\b)=\\
\sum_{l=\nq,\dots,n} 
\left(
\sum_{\l\in\O(l,n;\b)} \frac1{N^{\b n^2}}\,
\|P_{\l'}(z;\b)\|_N^2 \,\, |P_\l(\z_1,\dots,\z_n \at  \b;{\tsize \frac 1\b})|^2 
\right)
\endmultline \tag 5.3 
$$
and apply to each summand the asymptotic relation
\tht{4.9} with
$$
r=lp\,,\quad s=lq-n \,.
$$

Then since
$$
\fL(lp,lq-n;n,{\tsize \frac 1\b})=0
$$
the $l$-th piece of the sum \tht{5.3} becomes
a Riemann integral sum for the following integral 
$$
(lp)!\,(lq-n)! \, C^n_{lp,lq-n}(\b) \int\limits \Sb
1\ge x_1 \ge \dots \ge x_{lp} \ge 0 \\
\xi_1\ge \dots \ge \xi_{lq-n} \ge 0\endSb
dx\,d\xi \,\,\,
G(x\,|\,\xi;\b) \, 
\left|\Fh {x\,|\,\xi}{i y}{\tsize \frac1\b}\right|^2 \,.
$$
where
$$
\multline 
G(x_1,\dots,x_r\,|\,\xi_1,\dots,\xi_s;\b)= 
\\ 
\frac{V(x)^{\frac2\b}\, V(\xi)^{2\b}}{r!s!}
\prod_{i=1}^r \left(x_i(1-x_i)\right)^{\frac1\b-1}
\prod_{j=1}^s \left(\xi_i\left(1+\frac{\xi_i}\b\right)\right)^{\b-1} 
\prod_{i=1}^r \prod_{j=1}^s \left(x_i+\frac{\xi_i}\b\right)^{-2} \,.
\endmultline 
$$
Since the integrand is symmetric in $x_1,\dots,x_{lp}$
and also symmetric in $\xi_1,\dots,\xi_{lq-n}$ we can rewrite
the last integral as follows: 
$$
C^n_{lp,lq-n}(\b)
\int_0^1 dx_1 \dots \int_0^1 dx_{lp} \,
\int_0^\infty  d\xi_1 \dots \int_0^\infty d\xi_{lq-n} \,
G(x\,|\,\xi;\b) \, 
\left|\Fh {x\,|\,\xi}{i y}{\tsize \frac1\b}\right|^2 \,,
$$
which is exactly the $l$-th piece of the sum in \tht{1.2}.


\Refs

\widestnumber\key{LPS2}

\ref
\key C
\by F.~Calogero
\paper Solution of the one-dimensional $N$-body problem with
quadratic and/or inversely quadratic pair potential
\jour J.~Math.\ Phys.\
\vol 12 \pages 419-439 \yr 1971
\endref


\ref
\key D
\by C.~Dunkl
\paper Hankel transforms associated to finite reflection groups
\inbook Hypergeometric functions on domains of positivity,
Jack polynomials, and applications (Tampa, FL, 1991), 
Contemp.\ Math.\ 
\vol 138 \yr 1992
\pages 123--138
\endref







\ref
\key F1
\by P.~J.~Forrester
\paper Selberg correlation integrals and the $1/r\sp 2$ quantum
many-body system
\jour Nuclear Phys.~B
\vol 388 \yr 1992 \issue 3 \pages 671--699
\endref 

\ref
\key F2
\bysame
\paper
Addendum to: "Selberg correlation integrals and the $1/r\sp 2$
quantum many-body system" 
\jour  Nuclear Phys.~B \vol 416 \yr 1994
\issue 1 \pages 377--385
\endref 

\ref
\key F3
\bysame
\paper
Integration formulas and exact calculations in the
Calogero-Sutherland model
\jour Modern Phys.~Lett.~B
\vol 9 \yr 1995 \issue 6
\pages 359--371
\endref 

\ref 
\key Ha
\by Z.~N.~C.~Ha
\paper Fractional statistics in one dimension: view from
an exactly solvable model
\jour Nuclear Physics B [FS]
\vol 435 \yr 1995 \pages 604--636
\endref

\ref
\key J
\by  M.~F.~E.~de Jeu
\paper The Dunkl transform
\jour Invent.\ Math.\
\vol 113 \yr 1993 \issue 1 \pages 147--162
\endref

\ref
\key Ka
\by K.~Kadell
\paper An integral for the product of two Selberg-Jack symmetric polynomials.
\jour Compositio Math.\
\vol 87 
\yr 1993
\issue 1
\pages 5--43
\endref  

\ref
\key K
\by J.~Kaneko
\paper Selberg integrals and hypergeometric
functions associated with Jack polynomials
\jour SIAM J.\ Math.\ Anal.\ 
\vol 24 \yr 1993 \pages 1086--1110
\endref

\ref
\key Kn
\by F.~Knop
\paper Symmetric and non-Symmetric quantum Capelli
polynomials
\paperinfo to appear
\endref

\ref
\key KOO
\by S. Kerov, A.~Okounkov, and G.~Olshanski
\paper The boundary of Young graph with Jack edge
multiplicities
\paperinfo to appear, q-alg/9703037
\endref 

\ref
\key KS
\by F.~Knop and S.~Sahi
\paper Difference equations and symmetric polynomials
defined by their zeros
\jour Internat.\ Math.\ Res.\ Notices 
\yr 1996 \issue 10 \pages 473--486
\endref 



\ref
\key La
\by M.~Lassalle
\paper Une formule de bin\^ome
g\'en\'eralis\'ee pour les polyn\^omes de Jack
\jour Comptes Rendus
Acad.\ Sci.\ Paris, S\'er.\ I
\vol 310 \yr 1990
\pages 253--256
\endref



\ref
\key LPS1
\by  F.~Lesage, V.~Pasquier,  and D.~Serban
\paper  Single-particle Green function in the Calogero-Sutherland 
model for rational couplings $\beta=p/q$
\jour Nuclear Phys.\ B 
\vol 466 \yr 1996 \issue 3
\pages 499--512
\endref

\ref
\key LPS2
\bysame 
\paper 
Dynamical correlation functions in the Calogero-Sutherland model
\jour Nuclear Phys.\ B 
\vol 435 \yr 1995
\issue 3 \pages 585--603
\endref


\ref
\key M
\by I.~G.~Macdonald
\book Symmetric functions and Hall polynomials, 
second edition
\publ Oxford University Press \yr 1995
\endref

\ref 
\key Me
\by M.~L.~Mehta
\book Random matrices
\publ Academic Press
\publaddr New York
\yr 1991
\endref

\ref
\key Ok1
\by A.~Okounkov
\paper
Quantum immanants and higher Capelli identities
\jour Transformation Groups
\vol 1 \issue 1 \yr 1996 \pages 99-126
\endref



\ref
\key Ok2
\bysame
\paper
(Shifted) Macdonald polynomials: $q$-integral
representation and combinatorial formula
\paperinfo
to appear, q-alg/9605013
\endref



\ref
\key Ok3
\bysame
\paper
Shifted Macdonald polynomials with 3 parameters
 and binomial formula for Koornwinder polynomials
\paperinfo
to appear, q-alg/9611011
\endref

\ref
\key OO
\by A.~Okounkov and G.~Olshanski
\paper Shifted Schur functions
\jour Algebra i Analiz
\vol 9
\yr 1997
\pages No.~2
\lang Russian
\transl\nofrills English version to appear in St.~Petersburg Math. J. 
{\bf 9} (1998), No.~2
\endref


\ref
\key OO2
\bysame
\paper Shifted Jack polynomials, binomial formula,
and applications
\jour Math.\ Res.\ Letters
\vol 4 \yr 1997 \pages 69--78
\endref

\ref
\key Op
\by E.~M.~Opdam
\paper Dunkl operators, Bessel functions, and the
discriminant of a finite Coxeter group
\jour Compos.\ Math.\
\vol 85 \yr 1993
\pages 333-373
\endref

\ref 
\key OV
\by G.~Olshanski and A.~Vershik
\paper Ergodic unitarily invariant measures
 on the space of infinite Hermitian matrices
\inbook in Contemporary Mathematical Physics. F.~A.~Berezin's memorial volume
\bookinfo American Mathematical Society Translations, Series 2, Vol. 175 
(Advances in the Mathematical Sciences --- 31)
\eds R.~L.~Dobrushin, R.~A.~Minlos, M.~A.~Shubin, A.~M.~Vershik
\publ Amer. Math. Soc.
\publaddr Providence, R.I.
\yr 1996 \pages 137--175
\endref

\ref
\key S1
\by S.~Sahi
\paper The spectrum of certain invariant differential operators
associated to a Hermitian symmetric space
\inbook Lie theory and geometry: in honor of Bertram Kostant,
Progress in Mathematics
\vol 123
\eds J.-L.~Brylinski, R. Brylinski, V.~Guillemin, V. Kac
\publ Birkh\"auser
\publaddr Boston, Basel
\yr 1994
\endref


\ref
\key S2
\bysame
\paper Interpolation, integrality, and a generalization
of Macdonald's polynomials
\jour Internat.\ Math.\ Res.\  Notices 
\yr 1996 \issue 10 \pages 457--471
\endref


\ref
\key St
\by R.~P.~Stanley
\paper Some combinatorial properties of Jack symmetric functions
\jour Adv.\ in Math.\
\vol 77 \yr 1989 \pages 76--115
\endref

\ref
\key Su
\by B.~Sutherland
\paper Exact results for a quantum many-body problem in one
dimension
\jour Phys.\ Rev.\ \vol A4 \pages 2019--2021 \yr 1971
\moreref \jour Phys.\ Rev.\ \vol A5 \pages 1372--1372 \yr 1972
\endref

\ref
\key VK1
\by  A.~M.~Vershik and S.~V.~Kerov
\paper  
Asymptotic theory of 
characters of the infinite symmetric group
\jour Funct.\ Anal.\ Appl.\
\vol 15 \yr 1981 \pages 246--255
\endref

\ref 
\key VK2 
\bysame 
\paper Characters and factor representations of the
infinite unitary group  
\jour Soviet Math.\ Dokl.\
\vol 26 
\yr 1982 \pages 570--574
\endref

\endRefs
\enddocument

\end